\documentclass[10pt,aps,prb,twocolumn,showpacs,amssymb,floatfix]{revtex4-1}
\usepackage{amsmath,graphics,epsfig,mathrsfs}
\usepackage{epstopdf}
\usepackage{bm}
\usepackage{color}
\usepackage{soul}

\begin{document}
\title{Crossover between spin swapping and Hall effect in metallic systems}
\author{Hamed Ben Mohamed Saidaoui$^1$}
\author{Y. Otani$^{2,3}$}
\author{A. Manchon$^{1}$}\email{aurelien.manchon@kaust.edu.sa}
\affiliation{$^1$Physical Science and Engineering Division, King Abdullah University of Science and Technology (KAUST), Thuwal 23955-6900, Kingdom of Saudi Arabia\\
$^2$Institute for Solid State Physics, University of Tokyo, 5-1-5 Kashiwa-no-ha, Kashiwa, Chiba 277-8581, Japan\\
$^3$RIKEN-CEMS, 2-1 Hirosawa, Wako, Saitama 351-0198, Japan}

\begin{abstract}
We theoretically study  the crossover between spin Hall effect and spin swapping, a recently predicted phenomenon that consists in the interchange between 
the current flow and its spin polarization directions [Lifshits and D'yakonov, Phys. Rev. Lett. {\bf103}, 186601 (2009)]. Using a tight-binding model with spin-orbit coupled disorder, spin Hall effect, spin relaxation and spin swapping are treated on equal footing. We demonstrate that spin Hall effect and spin swapping present very different dependences as a function of the spin-orbit coupling and disorder strengths. As a consequence, we show that spin swapping may even exceed spin Hall effect. Three set-ups are proposed for the experimental observation of the spin swapping effect in metals.
\end{abstract}
\maketitle
\section{Introduction}
The spin-orbit coupling (SOC) is a relativistic effect that couples the particle's spin degree of freedom to its orbital angular momentum. In solid-state, this coupling originates from the interaction between the carrier's spin and the magnetic field that it experiences in its rest frame in the presence of a potential gradient (crystal field, defects etc.). The locking between spin and orbital angular momenta has spectacular consequences in normal metals such as, but not limited to, spin Hall effect \cite{dyakonov-perel1971,kato_science2004,wunderlich_prl2005}, spin galvanic effect \cite{ivchenko1989,ganichev2001,ganichev2002} and spin relaxation \cite{dyakoperel,fabian2007,wu2010}. In semiconductors\cite{dyakonov2008}, metals\cite{jungwirth2012,hoffman2013,sinova2014} and more recently topological insulators\cite{Hasan2010}, SOC is now commonly engineered and exploited to generate pure spin currents, thereby enabling the electrical manipulation of the spin degree of freedom in the absence of an external magnetic field.\par 

The most emblematic effect induced by SOC is probably the spin Hall effect (SHE) originally predicted by Dyakonov and Perel \cite{dyakonov-perel1971} and revived thirty years later by Hirsch \cite{hirsch} and Zhang \cite{zhang2000}. In analogy with the ordinary Hall effect, SHE consists in the generation of a transverse   
spin current in response to an electric field applied in the longitudinal direction. Geometrically, the spin current has the form $J_{j}^i=\epsilon_{ijk}\alpha_{\rm H}(\sigma_0/e) E_k$, where $J_{j}^i$ is a spin current polarized along the direction $i$ and flowing along the direction $j$, $E_k$ is the electric field applied along the direction $k$, $\sigma_0$ is the longitudinal conductivity, $\alpha_{\rm H}$ is the spin Hall angle (measured in \%) and $\epsilon_{ijk}$ is Kronecker symbol. In their pioneering theory, Dyakonov and Perel derived SHE from Mott scattering on spin-orbit coupled impurities, i.e. from the asymmetric spin-dependent scattering of an initially unpolarized electron flow \cite{smit_1955,smit_1958}. This mechanism is usually referred to as {\em extrinsic} since it depends upon the presence of impurities in the system. Another {\em extrinsic} effect that occurs upon scattering is the shift of the position of the incoming wave packet, called side jump scattering \cite{berger_1970}. This shift renormalizes the velocity operator by creating an effective anomalous velocity proportional to the number of impurities in the system \cite{lyo1972}.

Besides extrinsic contributions to SHE, the spin-orbit coupled band structure may also dramatically influence the velocity of traveling electron wave packets even in the absence of impurity scattering. Indeed the velocity operator, $\hat{\bm v}=\partial_{\bm p} \hat{H}$, is affected by the SOC present in the band structure which induces an additional Lorentz force that bends the trajectory of Bloch states \cite{karplus1954}. The effective magnetic field felt by the moving electron (also called Berry curvature\cite{berry,berryphase}) and arising from the band structure is {\em intrinsic} as it does not rely on impurity scattering. This contribution has been widely studied in the context of anomalous Hall effect \cite{Nagaosa2010}, for instance. The first prediction of impurity-free intrinsic SHE in non-centrosymmetry semiconductors \cite{sinova2004,murakami_science2003} paved the way towards the prediction and discovery of topological insulators\cite{Hasan2010}.\par

The search for efficient SHE has been quite intense in the past ten years \cite{jungwirth2012,hoffman2013,sinova2014}. After its pioneering observation in semiconductors \cite{kato_science2004,wunderlich_prl2005}, the attention has quickly drifted towards the exploration of metals \cite{valenzuela2006}. A wide variety of noble and transition metals has been investigated and while the exact values and the proper method to detect SHE is still under debate, clear indication of intrinsic SHE \cite{Guo2008,tanaka2008} has been established in 4$d$ and 5$d$ noble metals (Pt, Pd, Ta, W etc. - see Ref. \onlinecite{vila2007,mihajlovic_prl2009,mosendz2010,morota2011,pai2012,fujiwara_nature2013}), as well as in 3$d$ transition metals \cite{du2014}. Alternatively, it has been recently suggested that large spin Hall angle $\alpha_{\rm H}$ could be achieved by exploiting skew scattering on the resonant states of 5$d$ or 4$f$ impurities embedded in a light metal matrix, such as Cu or Al\cite{Guo2009,tanaka2009,fert2011,levy2013}. Evidence of large Hall angles from resonant skew scattering was reported in Pt- \cite{seki_nature2008,gu2010} and W-doped Au \cite{Laczkowski2014}, as well as Ir- \cite{niimi_prl2011} and Bi-doped Cu \cite{niimi_prl2012}. \par

Metals doped with heavy impurities present an interesting paradigm as they may be utilized to observe novel effects such as spin swapping (SSW), i.e. the conversion of a spin current $J_j^i$ into a spin current $J_i^j$ upon scattering by spin-orbit coupled impurities\cite{lifshits_prl2009,sadjina2012}. In the present work, we explore the nature of spin-orbit coupled transport within a disordered tight-binding model and demonstrate the emergence of SSW when the injected current is spin-polarized. In our model, SSW, SHE and spin relaxation all arise from the same spin-orbit coupled disorder which allows us to treat these different effects on equal footing. In particular, a crossover between SSW and SHE, controlled by the spin relaxation, is revealed and analyzed. This article is organized as follows. The physics of impurity-induced spin swapping is first outlined in Section \ref{s:ssw}. In Section \ref{s:tbm}, a single band tight-binding model and the corresponding non-equilibrium Green's function technique are presented. Section \ref{s:res} is devoted to the numerical results. Finally, Section \ref{s:obs}  suggests three setups for the experimental observation of SSW, and Section \ref{s:concl} concludes this study.

\section{Physics of spin swapping\label{s:ssw}}

\begin{figure}[h!]
  \centering
  \includegraphics[width=9cm]{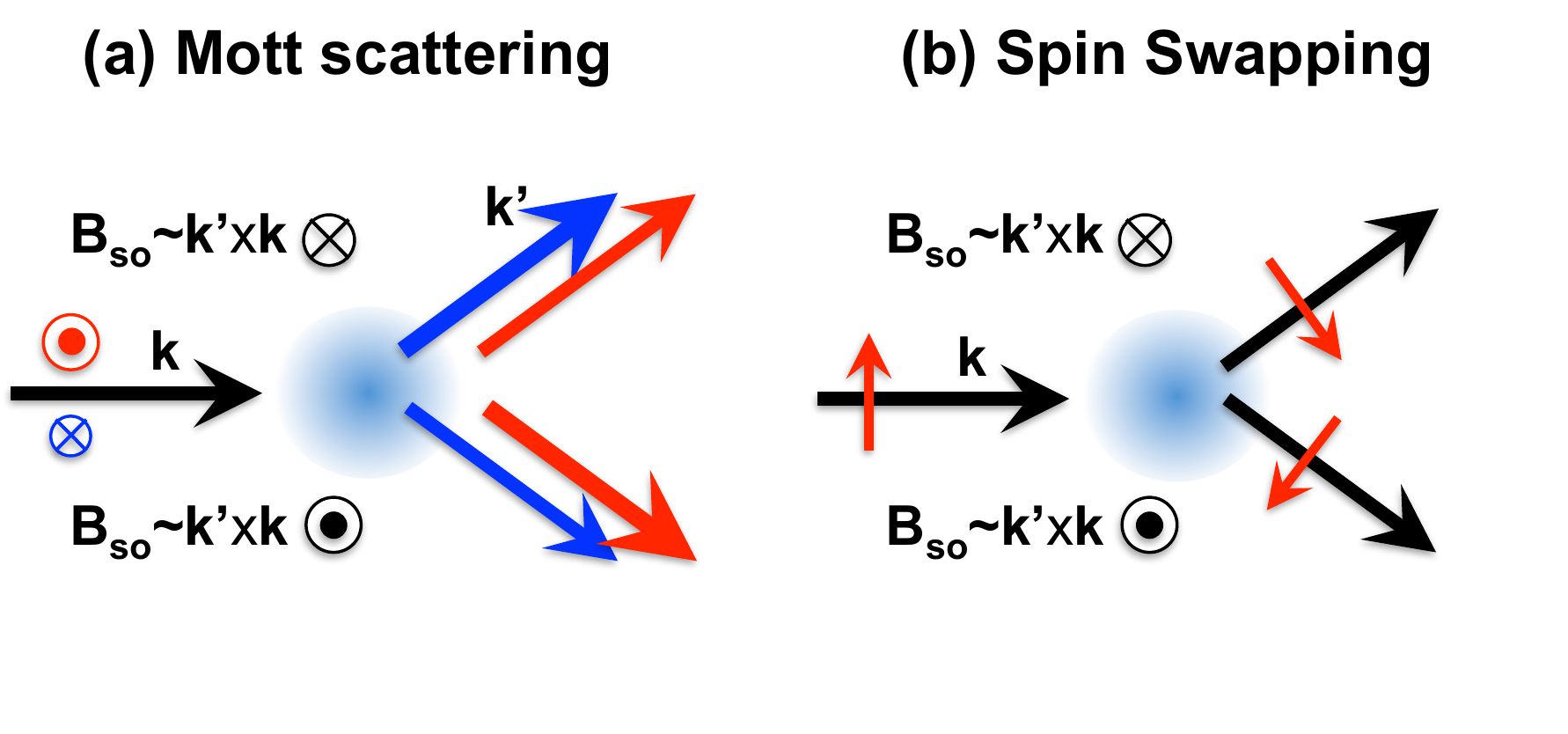}
  \caption{(Color online) (a) Sketch of Mott scattering by a spin-orbit coupled impurity. Electrons with an out-of-plane spin polarization pointing up have a larger probability to scatter towards the right while electrons with an out-of-plane spin polarization pointing down have a larger probability to scatter towards the left. (b) Sketch of spin swapping mediated by a spin-orbit coupled impurity. When the spin of the carrier lies in the scattering plane (${\bm k},{\bm k}'$), it experiences a magnetic field ${\bm B}_{\rm so}\propto{\bm k}'\times{\bm k}$ that depends on the direction towards which the spin is scattered. This induces a scattering-induced precession giving rise to the spin swapping. \label{fig:fig0}}
\end{figure}

The physics of extrinsic Mott skew scattering and SSW is illustrated on Fig. \ref{fig:fig0}(a) and (b), respectively. Let us first consider spin-dependent scattering against a spin-orbit coupled impurity. In real space, the impurity potential reads
\begin{equation}\label{eq:Himp}
\hat{H}^{\rm imp}=\sum_i V_{\rm imp}({\bm r}-{\bm r}_i)+(\lambda_{\rm so}/\hbar)\hat{\bm \sigma}\cdot({\bm\nabla} V_{\rm imp}({\bm r}-{\bm r}_i)\times{\hat{\bm p}}),
\end{equation}
where $V_{\rm imp}$ is the spin-independent impurity potential, and $\lambda_{\rm so}$ is the SOC parameter of the impurity. In the reciprocal space, this impurity potential becomes
\begin{equation}\label{eq:Hkk}
\hat{H}^{\rm imp}_{{\bm k}{\bm k}'}=\sum_iV^{\rm imp}_{{\bm k}{\bm k}'}e^{-i({\bm k}-{\bm k}')\cdot({\bm r}-{\bm r}_i)}[1+i\lambda_{\rm so}\hat{\bm \sigma}\cdot({\bm k}'\times{\bm k})].
\end{equation}
It is clear that the spin-orbit coupled part of the impurity potential acts like a magnetic field ${\bm B}_{\rm so}\propto {\bm k}'\times{\bm k}$ on the incoming electron spin $\hat{\bm \sigma}$, where ${\bm k}$ is the momentum of the incoming electron and ${\bm k}'$ is the momentum of the outgoing electron. Therefore, in the case of an unpolarized charge current, this magnetic field defines a local quantization axis so that one can phenomenologically separate electrons with a spin polarization pointing parallel and antiparallel to ${\bm B}_{\rm so}$, as illustrated on Fig. \ref{fig:fig0}(a). In this case, proper treatment of the collision integral leads to Mott scattering, i.e. electrons with a spin momentum pointing (anti)parallel to ${\bm k}'\times{\bm k}$ have the tendency to scatter towards the left (right). The resulting spin current is therefore polarized along ${\bm k}'\times{\bm k}$ and flows transversely to both ${\bm k}'\times{\bm k}$ and ${\bm k}$. This mechanism is at the core of the skew scattering mechanism giving rise to SHE as originally pointed out by Dyakonov and Perel \cite{dyakonov-perel1971}.\par

Let us now consider an incoming spin-polarized current (it can be either a pure spin current or accompanied by a charge flow), whose spin polarization lies in the scattering plane (${\bm k},{\bm k}'$). In this case, the incoming spin precesses around the effective magnetic field ${\bm B}_{\rm so}$. Since this magnetic field only exists upon scattering from ${\bm k}$ to ${\bm k}'$, the spin polarization of the outgoing current is re-oriented \cite{lifshits_prl2009}. Lifshits and Dyakonov tagged this effect "spin swapping" as the spin polarization and flow direction of the incoming spin current are swapped during this process: an incoming spin current $J_j^i$ gives rise to a spin current $J_i^j$ when $i\neq j$ (similarly a spin current, say, $J_x^x$ produces two spin currents $J_y^y$ and $J_z^z$). \par

Physically, SSW is nothing but a spin precession around the spin-orbit field ${\bm B}_{\rm so}$. Such precessions are well known in semiconductors possessing Rashba, Dresselhaus or Kohn-Luttinger SOC \cite{dyakonov2008}. This SOC-induced precession has been utilized in the Datta-Das transistor \cite{datta90,koo09} (see also Ref. \onlinecite{sadjina2012}) and its impact on spin transfer torque has been investigated in (Ga,Mn)As-based spin-valves \cite{haney2010} and single layers \cite{Li2015}. However, the impurity-induced spin swapping presents a number of interesting differences. First, in contrast with coherent spin precession around Rashba or Dresselhaus SOC, spin swapping arises from spin precession stemming from incoherent scattering. It is quite remarkable that this effect survives the disorder configurational average, as does extrinsic SHE. A spin diffusion equation in the presence of spin-orbit coupled impurities within the first Born approximation has been derived by Shchelushkin and Brataas \cite{Shchelushkin} a few years before the prediction of SSW by Lifshits and Dyakonov. The charge and spin current equations obtained in Ref. \onlinecite{Shchelushkin} are reproduced below
\begin{eqnarray}\label{eq:jcsw}
{\bm j}_c/\sigma_0&=&-{\bm \nabla}\mu_c+\frac{\xi_{\rm so}}{\lambda k_{\rm F}}{\bm\nabla}\times{\bm \mu}_s,\\\label{eq:jssw}
e{\bm J}_i/\sigma_0&=&-(1+\frac{2\xi_{\rm so}}{3})\nabla_i{\bm \mu}_s-\frac{\xi_{\rm so}}{\lambda k_{\rm F}}{\bm e}_i\times{\bm\nabla}\mu_c+\frac{2\xi_{\rm so}}{3}{\bm \nabla}\mu_s^i.\nonumber\\
\end{eqnarray}
Here, $\mu_c$ is the spin-independent chemical potential, $\mu_s^i$ is the $i$-th component of the spin-dependent chemical potential, and $\xi_{\rm so}=\lambda_{\rm so}k_{\rm F}^2$ is the unitless spin-orbit parameter. The metallic system is described in terms of its free electron Fermi wave vector $k_{\rm F}$ and its mean free path $\lambda$. ${\bm j}_c$ is the charge current density and ${\bm J}_i$ is the spin current density defined as a vector in the direction of the spin polarization, flowing along the direction ${\bm e_i}$. The first terms in Eqs. (\ref{eq:jcsw}) and (\ref{eq:jssw}) are the diffusion terms, the second terms ($\propto \frac{\xi_{\rm so}}{\lambda k_{\rm F}}$) are the side jump contribution producing SHE [Eq. (\ref{eq:jssw})] and inverse SHE  [Eq. (\ref{eq:jcsw})]. The third term in Eq. (\ref{eq:jssw}) is the spin swapping effect. Note that since these equations are derived within the first Born approximation (i.e. up to the second order in impurity potential $|V^{\rm imp}_{{\bm k}{\bm k}'}|^2$ only), skew scattering is neglected (see also Ref. \onlinecite{Shchelushkin2006}). The spin swapping term in Eq. (\ref{eq:jssw}) clearly converts ${\bm \nabla}\mu_s^i$ into ${\bm J}_i$, which is consistent with Lifshits and Dyakonov theory \cite{lifshits_prl2009}. A simplified version of the drift-diffusion equations, Eqs. (\ref{eq:jcsw})-(\ref{eq:jssw}), has been numerically investigated by Sadjina et al. \cite{sadjina2012}, neglecting SHE.\par

A second interesting aspect of SSW is that since it comes from impurity scattering, it can be engineered by using resonant scattering on heavy impurities embedded in a light metal host as suggested previously for SHE \cite{Guo2009,tanaka2009,fert2011,levy2013}. This aspect is beyond the scope of the present work, but deserves further investigations.\par

\section{Tight binding model\label{s:tbm}}

The drift-diffusion model presented in the previous section, Eqs. (\ref{eq:jcsw})-(\ref{eq:jssw}), has been obtained within the first Born approximation, i.e. only accounting for effects proportional to the impurity density and up to the second order in impurity potential \cite{Shchelushkin,Shchelushkin2006}. Therefore, configurational averaging is performed and skew scattering is neglected. In this model, the ratio  between SHE [second term in Eq. (\ref{eq:jssw})] and (side-jump only) SSW [third term in Eq. (\ref{eq:jssw})]  is $\propto 1/\lambda k_{\rm F}$.

The objective of the present work is to investigate the nature of spin-orbit coupled transport (i.e. SSW, SHE and spin relaxation) over a broad range of disorder strengths and accounting for all the relevant contributions to extrinsic SHE (side-jump, skew scattering etc.). To do so, we consider a tight-binding model on which spin-orbit coupled disorder is implemented. By computing the non-equilibrium local spin density that accumulates at the edges of the sample, one can evaluate the relative magnitude of SSW and SHE, identify its dependence as a function of the disorder strength and clarify the role of spin relaxation.\par 

\begin{figure}[h!]
  \centering
  \includegraphics[width=8.7cm]{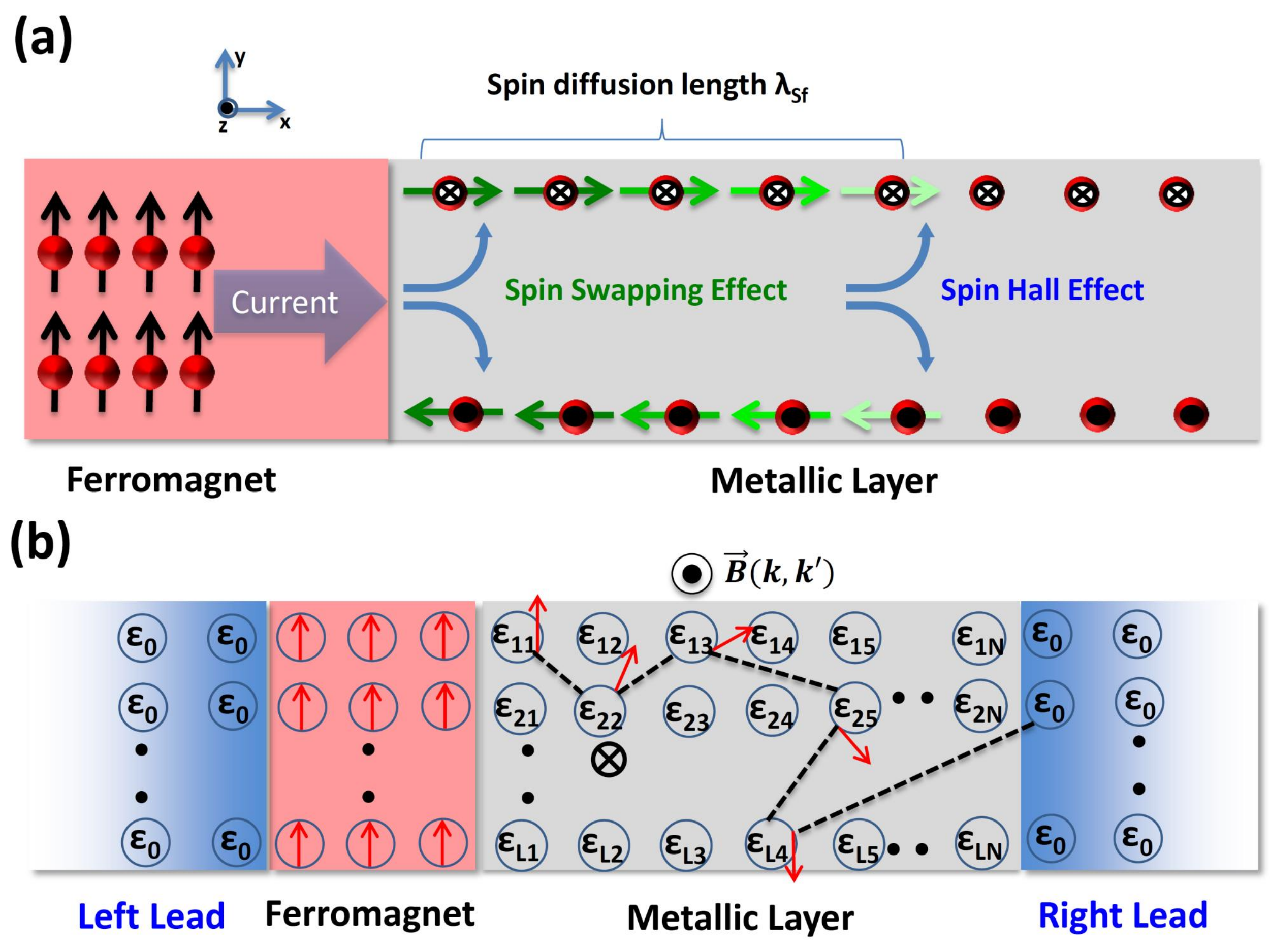}
  \caption{(Color online) (a) An illustration of the crossover between  SSW and SHE: SSW produces an edge spin accumulation that is polarized along the direction of injection (green arrows) and only survives close to the interface, while SHE builds up smoothly and results in a spin accumulation with a polarization normal to the plane. (b) Schematic of the tight binding model of the system. The circles refer to the atomic sites and $\epsilon$ stands for the on-site energy. The right and left leads are free from disorder with $\epsilon=\epsilon_0$. The central region is composed of a ferromagnet with polarization parallel to $\bm{y}$ and the disordered non-magnetic layer (symbolized by atoms having different on-site energies $\epsilon_{ij}=\epsilon_0+ \gamma_{ij}$, $\gamma_{ij}$ being the random onsite potential). The disorder-induced SOC is symbolized by the $\bm k$-dependent magnetic field ${\bm B}({\bm k},{\bm k}')$. The red arrows in the non-magnetic region stand for the carriers spins which we chose to have random directions to emphasize the effect of their coupling with the deflected momenta.}\label{fig:fig1}
\end{figure}

The metallic system we consider is sketched in Fig. \ref{fig:fig1}(a) and (b).  It is composed of a ferromagnet (FM) polarized along ${\bm y}$ and a normal metal (NM) with spin-orbit coupled disorder (see below). Following the discussion given in the previous section, we choose the FM polarization along ${{\bm y}}$ to insure that the spin density injected in the normal metal does not readily undergo SHE as long as the spin polarization is conserved [see Fig. \ref{fig:fig1}(a)]. Therefore, we expect SSW to take place in a region close to the interface and limited by the spin relaxation length. This SSW generates a spin accumulation at the edges of the sample whose polarization is aligned with the direction of injection ($\bm x$ in this example). While the injected spin accumulation is relaxed, over a distance of the order of the spin diffusion length $\lambda_{\rm sf}$, SHE smoothly takes over and a spin accumulation at the edges of the sample and polarized along the normal of the plane emerges. We numerically demonstrate this scenario using the tight-binding model described below.

The calculations carried out in the present work are based on the non-equilibrium Green's function technique implemented on a single band tight-binding model \cite{kwant}. All the  expectation values of the physical quantities of interest are given in real space. The whole system sketched in Fig. \ref{fig:fig1}(b), i.e. the FM/NM bilayer, is connected to two non-magnetic semi-infinite leads whose role boils down to maintaining the flow of the particles throughout the system and hence promoting the non-equilibrium regime. The total Hamiltonian of the central system reads
\begin{eqnarray}
{\hat H_s}& = &\sum_{i,j,\sigma,\sigma'}\{(\epsilon_{ij}\delta_{\sigma\sigma'}+\frac{\Delta_{ij}}{2}{\bm\Omega}_{ij}\cdot{\hat{\bm\sigma}}_{\sigma\sigma'}){\hat c}_{i,j,\sigma}^+ {\hat c}_{i,j,\sigma'}+h.c.\}\nonumber\\
&&-\sum_{i,j,\sigma}t({\hat c}_{i+1,j,\sigma}^+ {\hat c}_{i,j,\sigma}+{\hat c}_{i,j+1,\sigma}^+ {\hat c}_{i,j,\sigma}+h.c.) + {\hat H}_{so}.\nonumber\label{eq:H}
\\
\end{eqnarray}
Here the first term at the right-hand side of Eq. (\ref{eq:H}) is the spin-independent onsite energy in which $\epsilon_{ij}=\epsilon_0+\gamma_{ij}$, $\epsilon_0$ being the onsite energy constant and $\gamma_{ij}$ being a random onsite energy that introduces disorder in the system. The local disorder strength is randomly chosen such that $\gamma_{ij}\in[-\Gamma/2,\Gamma/2]$. The second term stands for the exchange interaction between the spin of the carriers and the local magnetic momentum of site ${(i,j)}$, $\Delta_{ij}$ is the exchange splitting, ${ {\bm \Omega}}_{ij}$ is the unit vector along the direction of the local magnetic moment on site ${(i,j)}$. In the present case, ${\bm \Omega}_{ij}={\bm y}$ in the FM layer and $\Delta_{ij}=0$ in the NM region. The operator ${\hat c}_{i,j,\sigma}^+$ (${\hat c}_{i,j,\sigma}$) creates (annihilates) a particle with spin $\sigma$ at position $(i,j)$. The third term in the Hamiltonian corresponds to the nearest neighbor hopping energy parameterized by the hopping integral $t$. \par

The SOC Hamiltonian ${\hat H}_{so}$ is given by the second term in Eq. (\ref{eq:Himp}) and can be discretized on the square lattice we consider. Since two spatial gradients are involved (${\bm\nabla}V_{\rm imp}$ and $\hat{\bm p}=-i\hbar{\bm\nabla}$), this term yields a spin-dependent next-nearest neighbor hopping contribution and can be rewritten \cite{pareek2001} 
\begin{eqnarray} 
{\hat H_{so}}=- i(\lambda_{\rm so}/a^2)\epsilon({\bm r}_a)({\bm r}_a\times{\bm r}_b)\cdot{\hat {\bm \sigma}},
\end{eqnarray}
where ${\bm r}_{a,b}$ are unit vectors pointing from the initial position to the nearest neighbor and from the nearest neighbor to the next-nearest neighbor, respectively, and $\epsilon({\bm r}_a)$ is the onsite energy of the nearest neighbor. Hence for a two-dimensional system, ${\hat H}_{so}$ reduces to two kinds of hopping involving the disordered-onsite energies\cite{pareek2001}
\begin{eqnarray}
&&t_{(i-1,j-1)\rightarrow (i,j)}=-i\alpha(\gamma_{i-1,j}-\gamma_{i,j-1})\hat{\sigma}_z,\\
&&t_{(i-1,j)\rightarrow (i,j-1)}=-i\alpha(\gamma_{i,j}-\gamma_{i-1,j-1})\hat{\sigma}_z,
\end{eqnarray}
where $\alpha=\lambda_{\rm so}/a^2$ and $a$ is the lattice parameter. The hopping from the site $(i-1,j-1)$ to the site $(i,j)$ can either pass by $(i,j-1)$ or by $(i-1,j)$ which give opposite contributions to the spin-orbit field. The same reasoning applies for the hopping from site $(i-1,j)$ to site $(i,j-1)$. In the present study, we choose the tight-binding parameters to be in units of the hopping energy $t$: the transport energy $E=t$, the exchange interaction $\Delta$ between the carrier spin and the local magnetic moment is, unless stated otherwise, given by $\Delta=t$. The SOC and disorder strengths ($\alpha$ and $\Gamma$) range from ($\alpha=0.2$ to $\alpha=0.6$ and $\Gamma=0.2\times t$ to $\Gamma=t$ respectively).\par

The effects we are interested in - SSW, SHE and spin relaxation - arise from the conjunction of disorder and SOC and hence only emerge when configurational averaging is properly carried out. The advantage of the tight-binding approach is that once the calculations are converged, mechanisms such as skew-scattering\cite{smit_1955,smit_1958} and side jump\cite{berger_1970} are fully accounted for. Therefore in the present work, each numerical result has been averaged over 10$^5$ configurations and convergency has been verified.

\section{Numerical results\label{s:res}}

\begin{figure}[ht]
  \centering
  \includegraphics[scale=0.42]{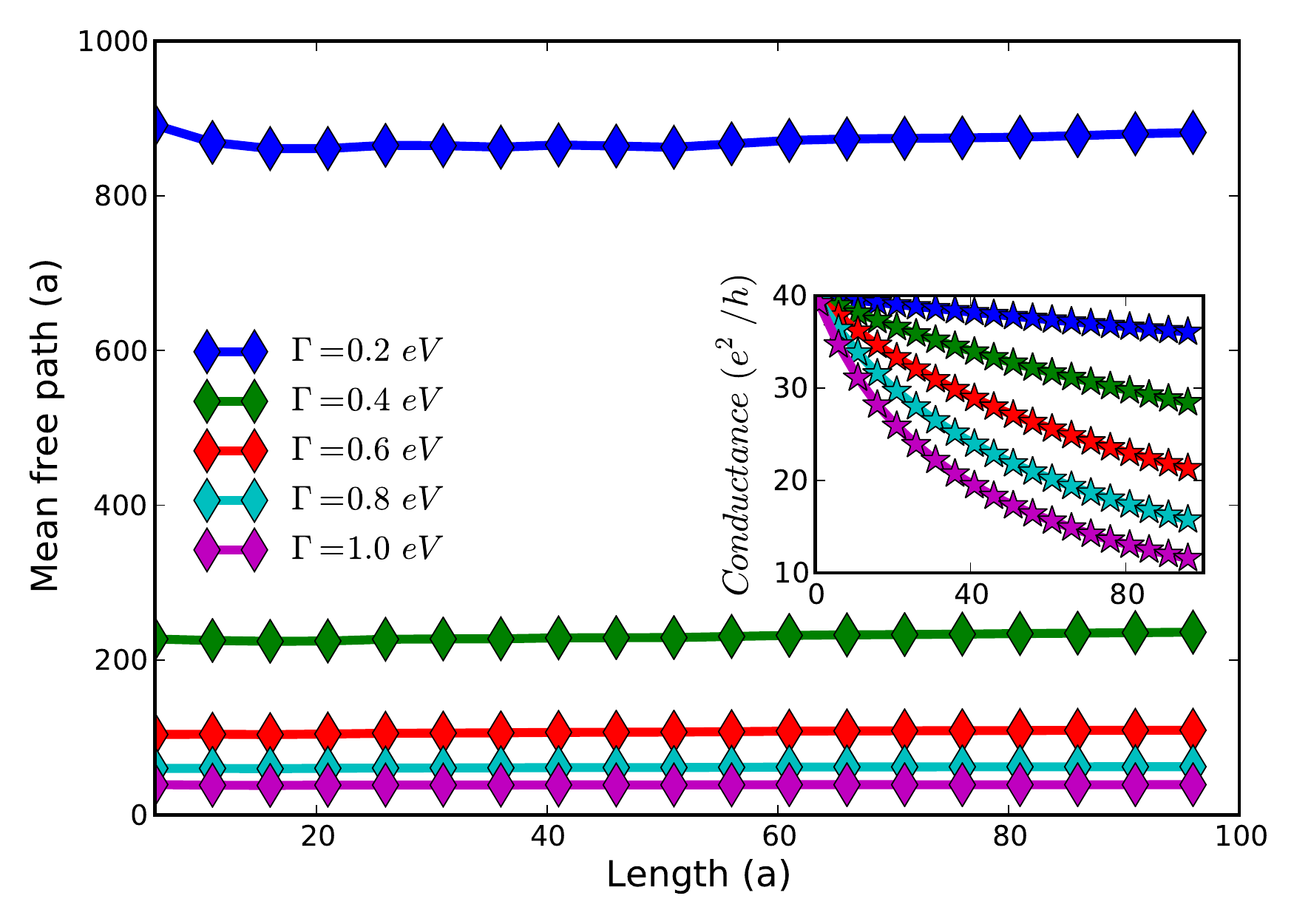}
  \caption{(Color online) Mean free path as a function of the sample length for different disorder strengths. The lower curves correspond to the normal (localization free) diffusive regime whereas a small divergence is obtained for the weaker disorder due to size effects (blue curve). The inset shows the corresponding conductances.\label{fig:fig2}}
\end{figure}

In this section we present the non-equilibrium calculations obtained using the Kwant code \cite{kwant}. An important criterion in the investigation of disorder-driven effects is the determination of the effective mean free path $\lambda$ of the system, as a function of the strength of the disorder. To do so, we used the semiclassical formula for the conductance
\begin{equation}\label{eq:cond}
G=G_0/\left(1+\frac{L}{\lambda}\right),
\end{equation}
where $G$ is the actual conductance of the sample, $G_0$ is the ballistic conductance, $L$ is the length of the sample and $\lambda$ is the mean free path. This formula indicates that as long as the mean free path is well defined, the ratio $\lambda_{\rm eff}\equiv L/(1-G_0/G)$ should remain constant when varying $L$. In principle, Eq. (\ref{eq:cond}) is only valid as long as  localization effects are weak and when SOC is absent. The effective mean free path extracted from our numerical calculations are displayed in Fig. \ref{fig:fig2} (the conductance is shown in the inset, for reference) and exhibits no dependence as a function of the length of the sample (except for very weak disorder - blue symbols - which indicates that the transport is not in the diffusive regime). These straight lines constitute a sane signature that the regime is diffusive in this parameter space. From now on, we choose the disorder strength in the range where the mean free path in Fig. \ref{fig:fig2} shows a constant profile.\par

\subsection{Crossover between spin swapping and spin Hall effects.}
\begin{figure}[h!]
  \centering
  \includegraphics[scale=0.52]{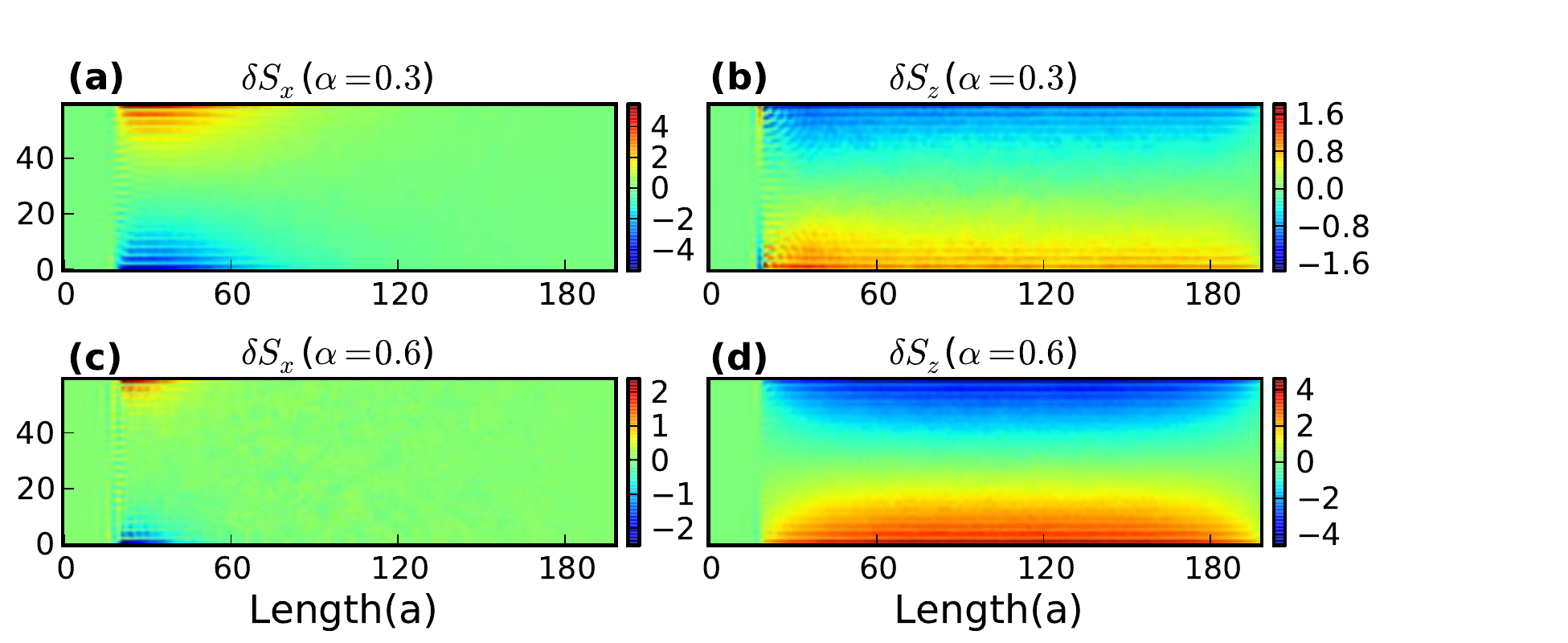}
  \caption{(Color online) Two dimensional mapping of the (a,c) $x$- and (b,d) $z$-components of the spin density (in unit of $10^{-3} \hbar/2$) demonstrating the crossover between SSW and SHE, for (a,b) $\alpha=0.3$ and (c,d) $\alpha=0.6$. The disorder strength is $\Gamma=1$eV.\label{fig:fig3}}
\end{figure}

\begin{figure}[h!]
  \centering
  \includegraphics[scale=0.35]{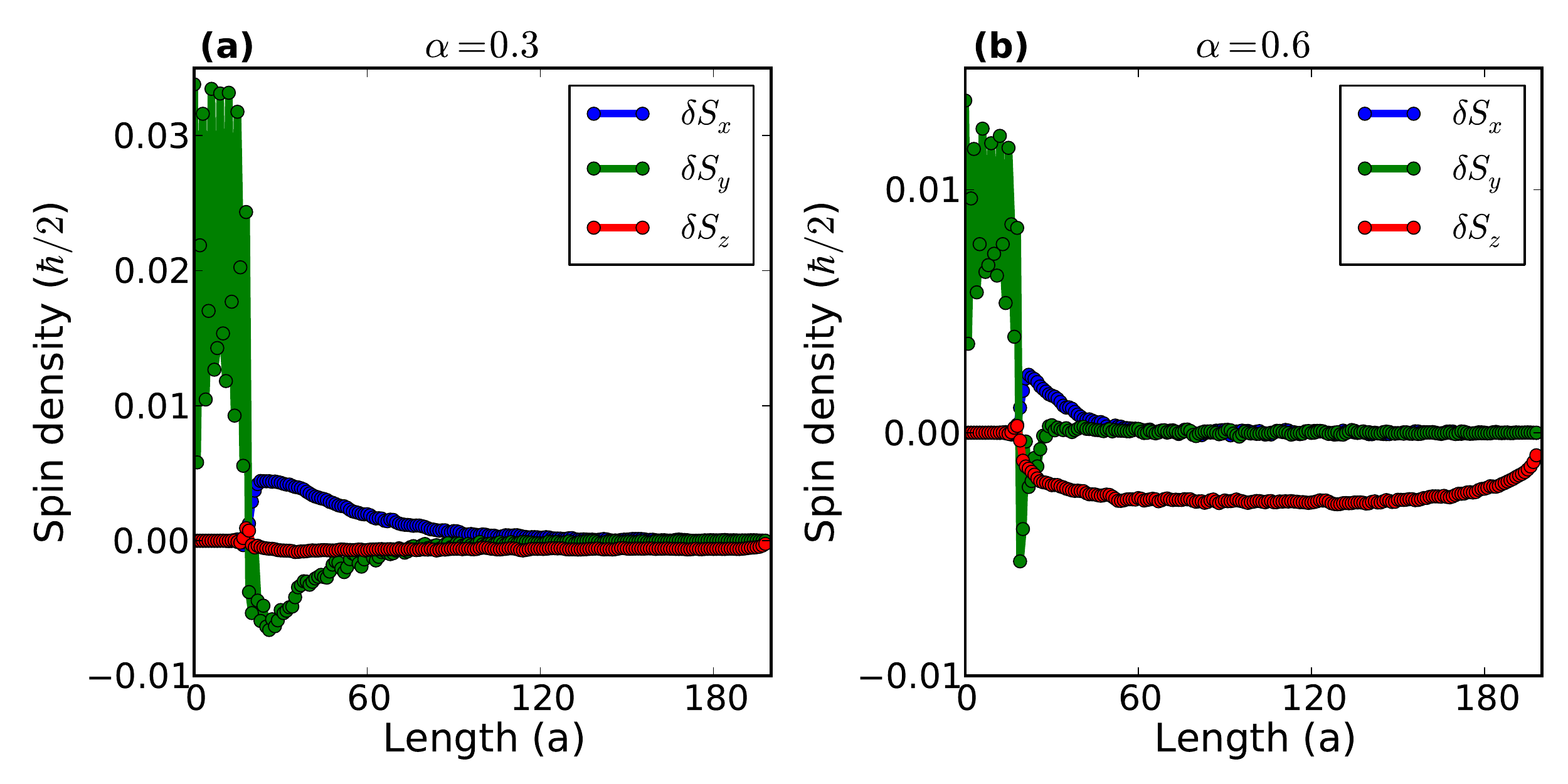}
  \caption{(Color online) Spatial profile of the ($x,y,z$)-components of the spin density calculated at the upper edge of the sample ($y=L/2$). The blue and red curves represent the spatial profile of the spin accumulation induced by SSW and SHE, respectively, while the green curve shows for the relaxation of the injected spin current. Note that the first 20 layers compose the ferromagnetic edge. The parameters are the same as in Fig. \ref{fig:fig3}. \label{fig:fig5}}
\end{figure}

In order to study the crossover between SHE and SSW, we consider the structure depicted in Fig. \ref{fig:fig1}(a). The current is injected from the FM layer into the NM layer along $x$-direction. Figure \ref{fig:fig3} displays the two dimensional map of the $x$- and $z$-components of the spin density, $S_x$ and $S_z$, for different strengths of the SOC unitless parameter, $\alpha$=0.3 and $\alpha$=0.6. The spatial profile of the three components of the spin density taken along a line passing by the center of the sample is shown in Fig. \ref{fig:fig5}. Since the magnetization of the FM layer is aligned along the $y$-direction, the injected current is spin polarized transversally to the direction of injection. The transverse component of the spin density $S_y$ simply relaxes in the NM due to SOC-driven spin relaxation [see Fig. \ref{fig:fig5}, green symbols]. Interestingly, the injected spin current $J_x^y$ immediately experiences spin swapping, which results in the accumulation of the spin density $S_x$ along the edges of the sample, close to the FM/NM interface. This accumulation vanishes over a length of the order of the spin diffusion length. Simultaneously, as the injected spin current gets progressively depolarized through spin relaxation, a spin density $S_z$ accumulates at the edges of the sample due to SHE. In contrast with $S_x$, $S_z$ survives away from the FM/NM and is maintained throughout the length of the sample. This crossover between SSW and SHE is therefore controlled by the spin relaxation length, in agreement with the scenario exposed in Section \ref{s:ssw}.\par

It clearly appears that the injected spin current $J_x^y$ is converted into $J_y^x$ upon scattering in the normal metal, in agreement with the phenomenological prediction of Lifshits and Dyakonov \cite{lifshits_prl2009} and with the drift-diffusion model given by Shchelushkin and Brataas \cite{Shchelushkin}. One can recover the spin density profile induced by SSW in Figs. \ref{fig:fig3}(a,c) by solving the diffusive transport equations, Eqs. (\ref{eq:jcsw})-(\ref{eq:jssw}), to the lowest order in SOC. In the configuration adopted in the present work, the spin accumulation injected in NM has the form $\mu^y_s=\mu_0 e^{-x/\lambda_{sf}}$, where $\mu_0$ depends on the material parameters \cite{valet1993} (conductivity, polarization, interfacial resistance etc.). In the presence of SOC, the magnitude $\mu_0$ acquires a dependence on $y$ and one finds that the spin accumulation at the edges is 
\begin{eqnarray}\label{eq:mus}
\mu_s^x|_L&=&(2\xi_{\rm so}/3)\mu_0\tanh(L/\lambda_{sf})e^{-x/\lambda_{sf}}.
\end{eqnarray}
This solution shows that the spin relaxation plays an important role in the build-up of the spin accumulation induced by SSW. In the absence of spin relaxation, no spin current is generated and therefore SSW is inactive. But on the other hand, when spin relaxation is too strong, obviously, the spin accumulation at the edges of the sample does not extend away from the interface. 

\subsection{Dependence on the SOC strength}

\begin{figure}[h!]
  \centering
  \includegraphics[scale=0.40]{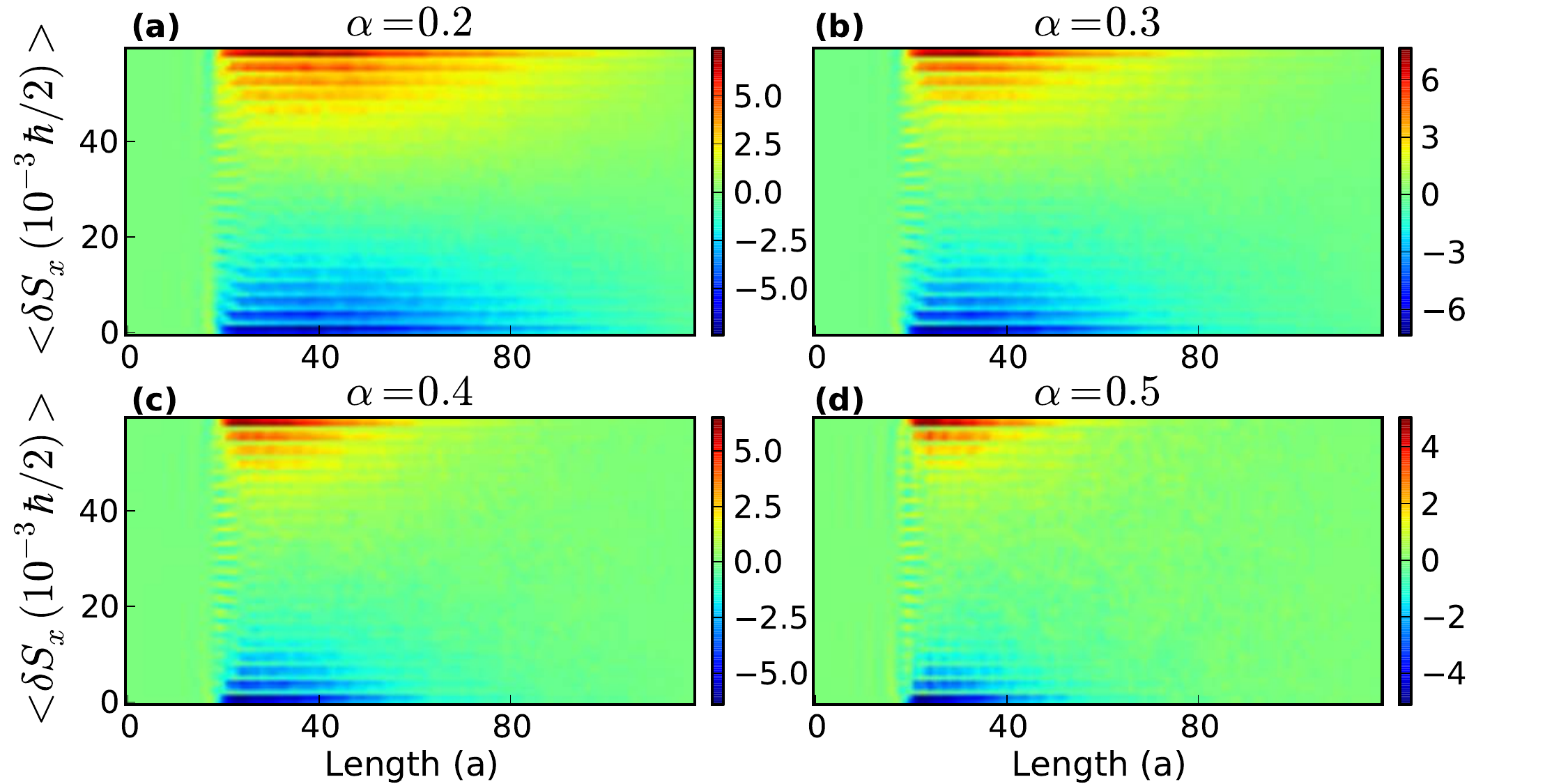}
  \caption{(Color online) Two dimensional mapping of the $x$-component of the spin density for various SOC strengths, ranging from $\alpha = 0.2$ to $\alpha = 0.5$. we choose $\Gamma=1$ eV in all the subplots.\label{fig:fig6}}
\end{figure}

The relative magnitude of the spin density accumulated through SSW and SHE depends on the strength of the SOC parameter in a nonlinear manner, which is at first sight quite surprising since both effects are, in principle, proportional to the SOC strength. However, the spin relaxation is detrimental to the spin swapping since it limits the amount of spin density available for the swapping. Indeed, Fig. \ref{fig:fig6} shows the two dimensional mapping of $S_x$ for different strengths of SOC. The induced spin density is localized at the edges of the sample and extents over $\lambda_{\rm sf}$ away from the interface. The spin relaxation length decreases when the SOC strength $\alpha$ increases from $\alpha=0.2$ ($\lambda_{\rm sf}\sim100 a$) to $\alpha=0.5$ ($\lambda_{\rm sf}\sim40 a$). \par

\begin{figure}[h!]
  \centering
  \includegraphics[scale=0.39]{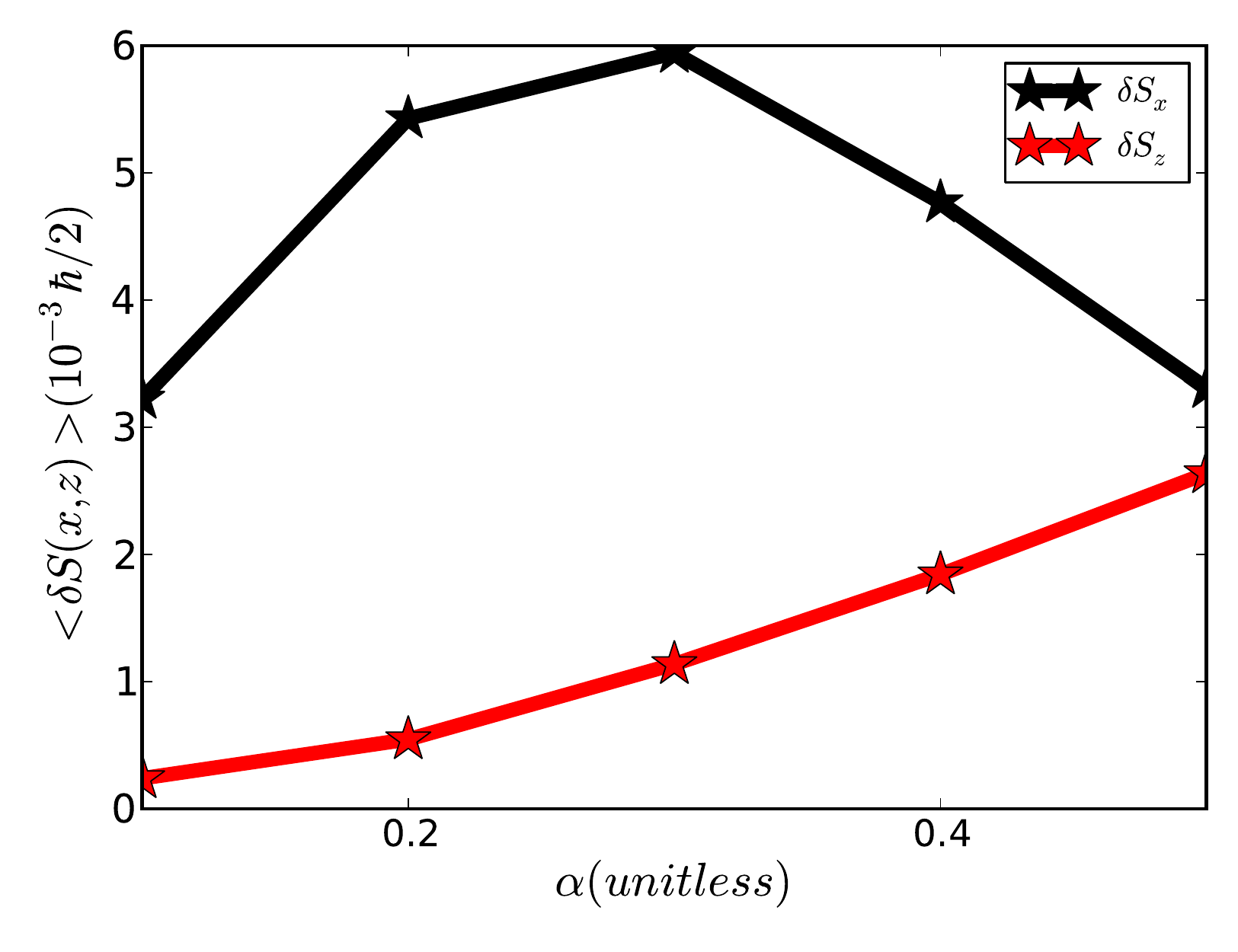}
  \caption{(Color online) $x$- and $z$-components of the spin density calculated for one chosen point at the upper edge of the sample with the coordinates ($x_0,L/2$). SSW-induced spin accumulation first increases and then decreases when increasing the SOC parameter $\alpha$, while SHE-induced spin accumulation monotonously increases. We have verified that the same behavior holds for different values of $x_0$. Here $\Gamma=1$ eV. \label{fig:fig4}}
\end{figure}

The interplay between the scattering events mentioned above in conjunction with the spin relaxation lead to the emergence of two regimes which characterize the dependence of the SSW accumulation on the SOC strength $\alpha$. The weak SOC regime corresponds to the enhancement of the spin swapping phenomena simply driven by the increase of the conversion efficiency with SOC\cite{lifshits_prl2009}, while the spin relaxation remains limited. For relatively larger values of SOC ($\alpha>0.3$), the SSW accumulation decreases in magnitude when $\alpha$ increases (Fig. \ref{fig:fig4}). The $x$-component of the spin density varies from $\sim 6\times10^{-3}\hbar /2$ to $\sim 4\times10^{-3}\hbar /2$ for $\alpha$ varying from $0.3$ to $0.5$. Indeed, while the conversion efficiency increases with $\alpha$, the spin relaxation reduces the amount of spin
current $J_x^y$ that is available for swapping. At large enough $\alpha$, the spin relaxation dominates and the spin swapping decreases. When $\alpha$ increases, the spin relaxation length (the extent of the spin accumulation at the edges) decreases from $\lambda_{sf}=\lambda_{sf}^0$ for $\alpha=0.2$ to $\lambda_{sf}=\lambda_{sf}^0/2$ for $\alpha=0.4$. It is remarkable to notice that while SSW has a non linear dependence as a function of $\alpha$, SHE homogeneously increases with $\alpha$. Indeed spin relaxation, as it depolarizes the injected spin current in NM, favors SHE.

\subsection{Influence of the disorder}

As mentioned in Section \ref{s:tbm}, the disorder is a mandatory ingredient in our model to get all the effects previously discussed, a fact that elucidates the pure extrinsic origin of the SSW effect. In order to identify the impact of disorder on SHE and SSW, we numerically calculate the spin density profile for various strengths of disorder (i.e. different mean free paths). Similarly to what has been noted for the dependence of the SSW accumulation on $\alpha$, the disorder dependence exhibits the two regimes mentioned above. While the SOC strength $\alpha$ controls the coupling between the spin and orbital momenta, the disorder strength $\Gamma$ controls the scattering probability off the impurities.\par
\begin{figure}[h!]
  \centering
  \includegraphics[scale=0.4]{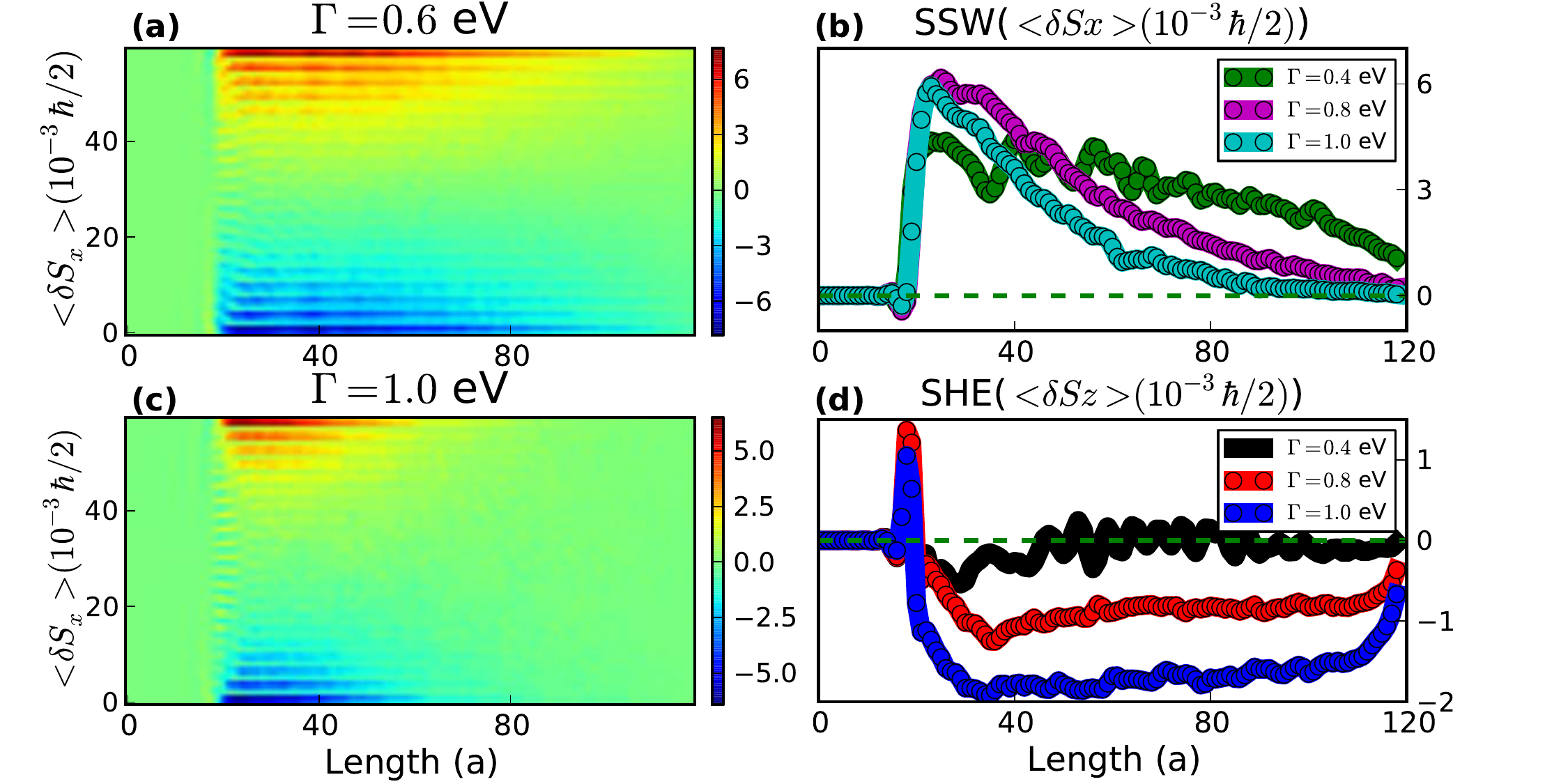}
  \caption{(Color online) Left panels: Two dimensional mapping of the $x$-component of the spin density for (a) $\Gamma=0.6$eV and (c) $\Gamma=1$eV; Right panels: Spatial profile of the (b) $x$- and (d) $z$-components of the spin density calculated at the upper edge of the sample ($y=L/2$), for different disorder strengths ($\Gamma$ =0.6, 0.8 and 1 eV). Here, we set $\alpha=0.4$. \label{fig:fig7}}
\end{figure}
Figure \ref{fig:fig7} represents the spatial profile of the $x$- and $z$- components of the spin accumulation at the edge of the sample for different disorder strengths. In these calculations, the effective mean free path ranges from $\lambda=100a$ ($\Gamma=0.4$ eV) to $\lambda=25a$ ($\Gamma=1$ eV). Quite remarkably, while the extension of the SSW-induced spin accumulation within the normal metal decreases when increasing the disorder strength [see Figs. \ref{fig:fig7}(a) and (c)], the overall magnitude of the edge spin accumulation is only weakly sensitive to $\Gamma$ [see Figs. \ref{fig:fig7}(b)]. This indicates that the magnitude of the SSW effect is mostly controlled by the spin relaxation strength, while the SSW efficiency itself remains independent on the disorder. This observation is consistent with the diffusive model which states that the SSW-induced spin current is proportional to the SOC strength only [see last term in Eq. (\ref{eq:jssw})], while the edge spin accumulation is controlled by the spin diffusion length [see Eq. (\ref{eq:mus})]. This behavior sharply contrasts with the dependence on the SHE-induced edge spin accumulation displayed in Fig. \ref{fig:fig7}(d). Indeed, our calculations show that the magnitude of the SHE-induced spin accumulation is on contrary very sensitive to the disorder strength and almost vanishes for $\Gamma<0.4$eV, whereas SSW survives. This is, again, consistent with the diffusive model, Eq. (\ref{eq:jssw}), that shows that in our model extrinsic SHE necessitates a large amount of disorder to emerge.

\section{Observing spin swapping\label{s:obs}}

\begin{figure}[h!]
  \centering
  \includegraphics[scale=0.4]{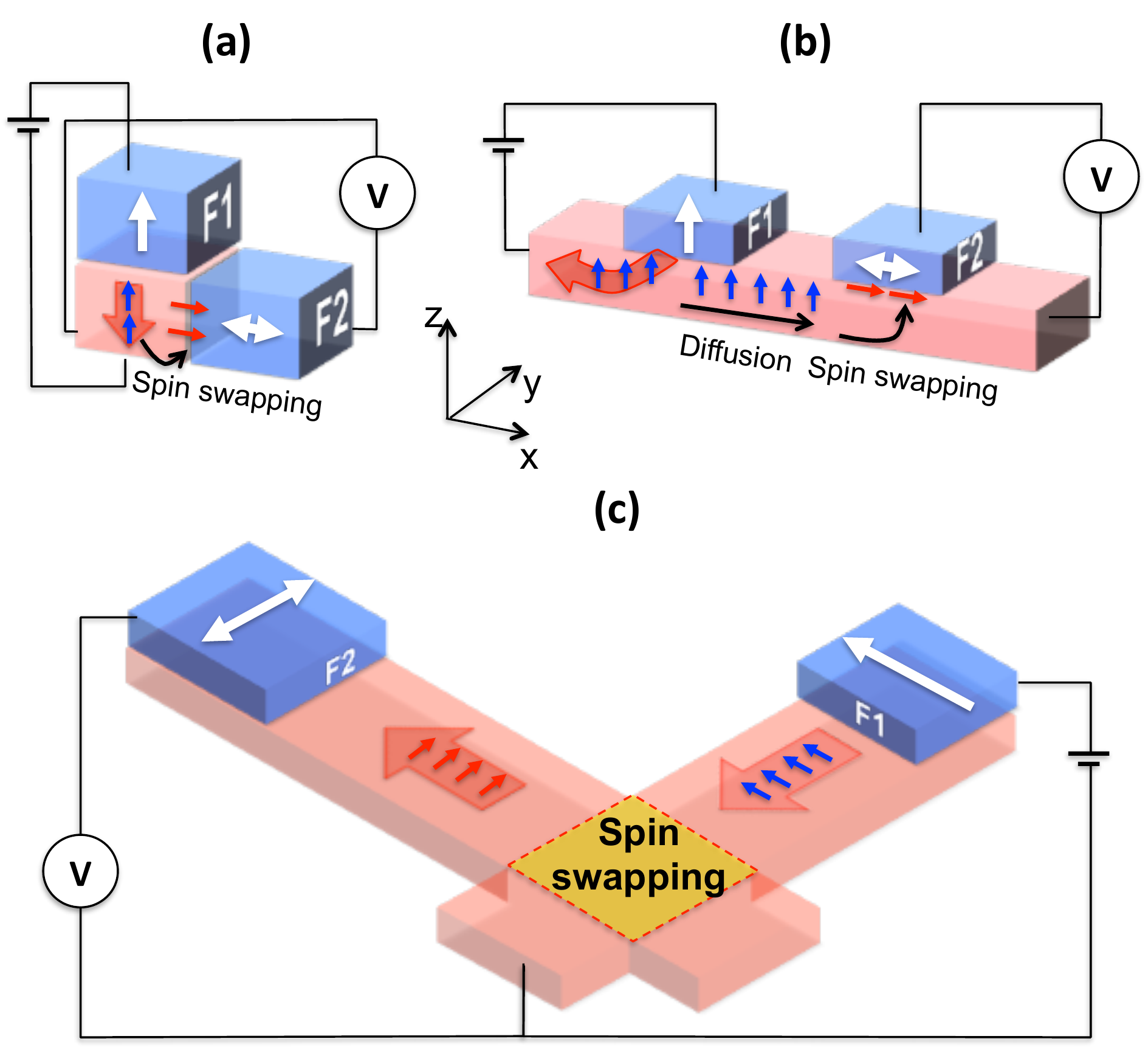}
  \caption{(Color online) Sketches of various setups for measuring spin swapping: (a) direct injection using a three terminal devices, (b) non-local injection in a one-dimensional "spin transistor" configuration and (c) direct injection in a Hall cross. The blue layers represent ferromagnets and the red layer represents a normal metal. The thick red arrow denotes the direction of injection while the small blue (red) arrows indicate the direction of the electron spins before (after) spin swapping.\label{fig:fig9}}
\end{figure}

Before concluding this study, we propose three setups for the experimental detection of the spin swapping. The first setup, displayed on Fig. \ref{fig:fig9}(a), relies on direct spin injection. A spin-polarized charge current is injected along the direction $\bm z$ from a polarizer (denoted F1) into a normal metal possessing strong SOC (red layer). Since the magnetization of the polarizer F1 is aligned along $\bm z$, the spin swapping generates a spin current flowing along the direction $\bm x$ and spin-polarized along $\bm x$. This spin current accumulates spins (red arrows) at the interface with a second ferromagnet, F2. By sweeping the magnetization of F2 along $\pm{\bm x}$ (or by simply reversing the direction of the injected current), one should collect a magnetoresistive signal arising from the interfacial spin accumulation produced by the spin swapping. The main constraint of this setup is that the area of the ferromagnet F2 in contact with the normal metal must be at most of the order of $\lambda_{\rm sf}^2$ to maximize the detection.\par

The second method we propose, displayed on Fig. \ref{fig:fig9}(b), relies on non-local spin injection and may be simpler to fabricate. A spin-polarized charge current is injected along $-{\bm x}$, from a polarizer (denoted F1) into the normal metal (red layer). A pure spin current, free of charge and spin-polarized along the magnetization direction of F1 aligned along $\bm z$, diffuses along $+{\bm x}$ (blue arrows). Through spin swapping, a spin current polarized along $\bm x$ and flowing along $\bm z$ is injected into a second ferromagnet F2 placed on top of the normal metal. Again, by sweeping the magnetization of F2 along $\pm{\bm x}$ or by reversing the direction of the injected current, one should collect a magnetoresistive signal arising from the spin swapping. The main constraint of this setup is that the distance between F1 and F2 must be smaller than $\lambda_{\rm sf}$. This setup presents some similarities with the famous spin field-effect transistor proposed by Datta and Das \cite{datta90}, whose operation relies on coherent spin precession induced by Rashba SOC rather than spin-orbit coupled impurities.\par

The two previous setups present the inconvenience of being limited by the spin diffusion length of the spin-orbit coupled metal. One alternative situation is depicted in Fig. \ref{fig:fig9}(c), which is a variant of the setup displayed in Fig. \ref{fig:fig9}(a). This structure consists in a Hall cross where a spin current is injected from a first ferromagnet F1 into the cross and spin-orbit coupled impurities are only present in the center of the cross. For instance, if one imagines a Hall cross made of Cu, doping the center of the cross with, say, Bismuth impurities would allow for the spin swapping to take place at this location only while preventing spin relaxation in the branches of the cross. The detection of the spin swapping-induced spin current can be achieved by performing a magnetoresistance measurement on F2. Another advantage is that the magnetization direction of F1 and F2 are both in-plane.

\section{Conclusion\label{s:concl}}
We have shown that spin-orbit coupled disorder induces a strong SSW that may even exceed extrinsic SHE in metals. The crossover between these two effects is controlled by the spin relaxation of the injected spin current. While both SSW and SHE emerge through scattering on spin-orbit coupled disorder, these two effects present a noticeably different dependence as a function of both the SOC and disorder strengths. Indeed, while SHE monotonously increases with the SOC strength, SSW displays a more complex behavior due to the competition with spin relaxation. In addition, while SHE is very sensitive to the disorder strength, it turns out that the SSW efficiency is rather controlled by spin relaxation than by the disorder strength itself.\par
Finally, we propose three experimental setups that shall allow for the experimental detection of the spin swapping in metals possessing impurities-driven spin-orbit coupling. Materials in which extrinsic SHE dominates over the intrinsic contributions, such as heavy metal-doped light metals such Au(Pt) \cite{Guo2009}, Au(W) \cite{Laczkowski2014}, Cu(Ir) \cite{niimi_prl2011}, Cu(Bi) \cite{niimi_prl2012}, or even rare-earth doped metals \cite{tanaka2009}, are attractive candidates for the observation of this effect as they associate strong spin Hall angle with a reasonably long spin diffusion length.

\acknowledgements
A.M. acknowledges inspiring discussions with T. Valet on the physics of spin swapping and H.B.M.S. thanks S. Feki for his valuable technical support. For computer time, this research used the resources of the Supercomputing Laboratory at King Abdullah University of Science and Technology (KAUST) in Thuwal, Saudi Arabia.
Y.O. acknowledges support from Grant-in-Aid for Scientific Research on Innovative Areas "Nano Spin Conversion Science" (Grant No. 26103002).

\end{document}